\documentclass[12pt]{article}
\usepackage{graphicx}
\usepackage{epstopdf}
\usepackage{epsfig}
\usepackage{bbm}

\newcommand{\be}{\begin{equation}}
\newcommand{\ee}{\end{equation}}

\def\eqref#1{(\ref{#1})}

\def\bsg{\ifmmode B\to X_s\gamma\else $B\to X_s\gamma$\fi}
\def\bsll{\ifmmode B\to X_s\ell^+\ell^-\else $B\to X_s\ell^+\ell^-$\fi}
\def\shat{\ifmmode \hat{s}\else $\hat{s}$\fi}
\def\OMITCOMPLETELY#1{}

\newcommand{\newc}{\newcommand}

\newc{\gsim}{\lower.7ex\hbox{$\;\stackrel{\textstyle>}{\sim}\;$}}
\newc{\lsim}{\lower.7ex\hbox{$\;\stackrel{\textstyle<}{\sim}\;$}}
\newc{\ie}{{\it i.e.}}
\newc{\etal}{{\it et al.}}
\newc{\mev}{\hbox{\rm\,MeV}}
\newc{\gev}{\hbox{\rm\,GeV}}
\newc{\tev}{\hbox{\rm\,TeV}}
\newc{\xpb}{\hbox{\rm\, pb}}
\newc{\xfb}{\hbox{\rm\, fb}}

\newc{\mtop}{m_t}
\newc{\mbot}{m_b}
\newc{\mz}{M_Z}
\newc{\mw}{M_W}
\newc{\alphasmz}{\alpha_s(M_Z)}
\newc{\swsq}{\sin^2\theta_W}
\newc{\cwsq}{\cos^2\theta_W}
\newc{\tw}{\tan\theta_W}
\newc{\cw}{\cos\theta_W}
\newc{\sw}{\sin\theta_W}
\newc{\BR}{\hbox{\rm BR}}
\newc{\zbb}{Z\to b\bar}
\newc{\Gb}{\Gamma (Z\to b\bar b)}
\newc{\Gh}{\Gamma (Z\to \hbox{\rm hadrons})}
\newc{\sgn}{\mbox{sgn}}

\newlength{\myem}
\newcounter{mysubequation}[equation]

\def\beq{\begin{equation}}
\def\eeq{\end{equation}}
\def\bea{\begin{eqnarray}}
\def\eea{\end{eqnarray}}
\def\slashchar#1{\setbox0=\hbox{$#1$}           
   \dimen0=\wd0                                 
   \setbox1=\hbox{/} \dimen1=\wd1               
   \ifdim\dimen0>\dimen1                        
      \rlap{\hbox to \dimen0{\hfil/\hfil}}      
      #1                                        
   \else                                        
      \rlap{\hbox to \dimen1{\hfil$#1$\hfil}}   
      /                                         
   \fi}                                         %
\catcode`@=11
\long\def\@caption#1[#2]#3{\par\addcontentsline{\csname
  ext@#1\endcsname}{#1}{\protect\numberline{\csname
  the#1\endcsname}{\ignorespaces #2}}\begingroup
    \small
    \@parboxrestore
    \@makecaption{\csname fnum@#1\endcsname}{\ignorespaces #3}\par
  \endgroup}
\catcode`@=12

\begin{document}

\baselineskip=18pt

\setcounter{footnote}{0}
\setcounter{figure}{0}
\setcounter{table}{0}

\begin{titlepage}
\begin{flushright}
HUTP-05/A0057\qquad\,\\
hep-th/0601001\qquad\,\\
\end{flushright}
\vspace{.0in}
\begin{center}
{\Large \bf  The String Landscape,}

{\Large \bf  Black Holes and}

{\Large \bf  Gravity as the Weakest Force}

\vspace{0.5cm}

{\bf Nima Arkani-Hamed, Lubo\v{s} Motl,\\
Alberto Nicolis and Cumrun Vafa}

\vspace{.5cm}

{\bf E-mail}: {\tt arkani, motl, nicolis, vafa (at) physics.harvard.edu}

\vspace{.5cm}

{\it Jefferson Laboratory of Physics, Harvard University,\\
Cambridge, Massachusetts 02138, USA}

\end{center}
\vspace{1cm}

\begin{abstract}
\medskip

We conjecture a general upper bound on the strength
of gravity relative to gauge forces in quantum gravity.
This implies, in particular,
that in a four-dimensional theory with gravity
and a $U(1)$ gauge field with gauge coupling $g$,
there is a new ultraviolet scale $\Lambda=g M_{\rm Pl}$,
invisible to the low-energy effective field theorist, which
sets a cutoff on the validity of the effective theory.
Moreover, there is some light charged particle with mass smaller than
or equal to $\Lambda$.  The bound is motivated by arguments
involving holography and absence of remnants, the (in) stability
of black holes as well as the non-existence of global
symmetries in string theory.  A sharp form of the conjecture
is that there are always light ``elementary'' electric and
magnetic objects with a mass/charge ratio smaller than
the corresponding ratio for macroscopic extremal
black holes, allowing extremal black holes to decay.
This conjecture is supported by a number of non-trivial examples in string
theory. It implies the necessary presence of new physics beneath the
Planck scale, not far from the GUT scale, and explains why some apparently
natural models of inflation resist an embedding in string theory.

\end{abstract}

\bigskip
\bigskip


\end{titlepage}


\tableofcontents
\vfill\eject

\section{Introduction}

By now it is clear that consistent theories
of quantum gravity can be constructed in the context
of string theory.  This can also be done in diverse dimensions
 by considering suitable compactifications.  This diversity, impressive
as it may be for a consistent theory to possess, poses a dilemma:
The theory appears to be more permissive than desired! However it
was recently suggested \cite{swamp} that the landscape of consistent
theories of gravity one obtains in string theory is by far smaller
than would have been anticipated by considerations of semiclassical
consistency of the theory.  The space of consistent low-energy
effective theories which cannot be completed to a full theory was
dubbed the `swampland'. Certain criteria were studied in
\cite{swamp} to distinguish the string landscape from the swampland.
For example, one such criterion was the finiteness of the number of
massless fields (see also \cite{doug} for a discussion of this
point).

In this paper we propose a new criterion which distinguishes parts
of the swampland from the string landscape.  This involves the
simple observation, clearly true in our own world,  that ``gravity
is the weakest force''. We promote this to a principle and in fact
find that surprisingly it is demanded by all consistent string
theory compactifications! Roughly speaking this is the statement
that there exist two ``elementary'' charged objects for which the
repulsive gauge force exceeds the attractive force from gravity.
More precisely, the conjecture we make is that this is
true for a stable charged particle which minimizes the ratio
$|M/Q|$.  In other words one of our main conjectures in this paper
is that (in suitable units) the minimum of $|M/Q|$ is less than 1.

We motivate this conjecture from various viewpoints.  In particular
we show how this conjecture follows from the assumption of
finiteness of the number of stable particles which are not protected
by a symmetry principle. This finiteness criterion nicely extends
some of the finiteness criteria discussed in \cite{swamp, doug} in
the context of bounds on the number of massless particles in a
gravitational theory. We show why if our conjecture were not true,
there would be an infinite tower of stable charged particles not
protected by any symmetry principle. The conjecture also ties in
nicely with the absence of global symmetries in a consistent theory
of gravity.

Our conjecture, if true, has a number of consequences:  For extremal
(not necessarily supersymmetric) black holes the bound is $M/|Q|=1$.
This suggests that there are corrections to this formula for smaller
charges, which makes it (in the generic case) an inequality
$M/|Q|<1$. Another aspect of our conjecture is that it naturally
suggests that the $U(1)$ effective gauge theory breaks down at a
scale $\Lambda$ well below the Planck scale $\Lambda \sim g M_{\rm Pl}$
(more precisely $\Lambda \sim \sqrt{\alpha/G_{\rm N}}$), where $g$ is the
$U(1)$ gauge coupling constant. These restrictions of low cutoff
scales and forced presence of light charged particles are very
surprising to the effective field theorist, who would not suspect
the existence of the new UV scale $\Lambda$. As long as the Landau
pole of the $U(1)$ is above the Planck scale, the low-energy
theorist would think that the cutoff of the effective theory should
be near $M_{\rm Pl}$, and if anything, smaller $g$ seems to imply that
the theory is getting even more weakly coupled.

Our conjecture, if true, has a number of consequences. If $g$ is chosen to 
be one of the Standard Model gauge couplings near the
unification scale, the scale $\Lambda$ is necessarily beneath the
Planck scale, close to the familiar heterotic string scale $\sim
10^{17}$ GeV. Furthermore, the observation of any tiny gauge
coupling, for instance in sub-millimeter tests of gravity, would
necessitate a low-scale $\Lambda$ far beneath the Planck/GUT
scales. Finally our conjecture also explains why certain classes of
apparently natural effective theories for inflation, involving
periodic scalars or axions with parametrically large, super-Planckian
decay constants, have resisted an embedding in string theory. \cite{bdine}.

This restriction has a number of phenomenological consequences. It
implies that, extrapolating the Standard Model to high energies,
there {\it must} be new scale $\Lambda$ beneath the Planck scale
with  $\Lambda \sim \sqrt{\alpha_{\rm GUT}/G_{\rm N}} \sim 10^{17}$ GeV,
and that any experimental observation of extremely weak gauge coupling
must be accompanied by new ultraviolet physics at scales far beneath
the Planck/GUT scales. 

\vspace{4mm}

\qquad\qquad\qquad\qquad\quad
\epsfig{file=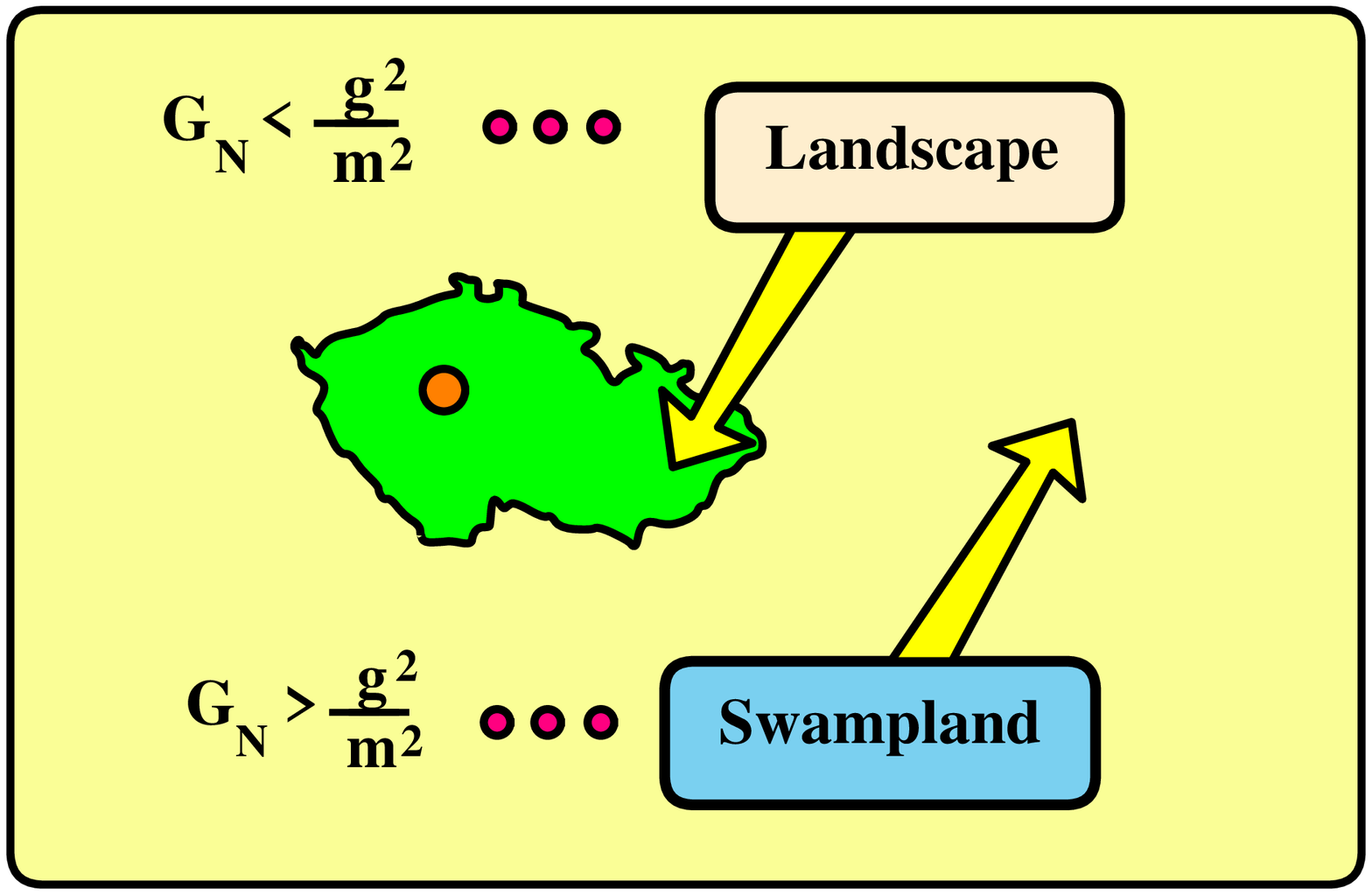,width=8cm}

\begin{quotation}{\bf Figure 1.} Consistent theories of quantum
gravity (the landscape) represent a small portion of the effective
field theories (the swampland) where additional conditions such as the
weakness of gravity are satisfied.
\end{quotation}

The organization of this paper is as follows:  In section 2 we
present some loose conjectures which motivate our more precise
conjectures discussed in section 3. In section 4 we present evidence
for our conjecture drawn from string theory. We conclude with a
discussion of certain additional points in section 5.

\section{Loose conjectures}

\subsection{Weak electric and magnetic gauge couplings}

Consider a 4-dimensional theory with gravity and a $U(1)$ gauge
field with gauge coupling $g$. Naively, the effective theory breaks
down near the scale $M_{\rm Pl}$ where gravity becomes strongly coupled,
and certainly nothing seems to prevent taking $g$ as small as we
wish; if anything the effective theory seems to become even more
weakly coupled and consistent.

Nonetheless, we claim that for small $g$ there is a hidden new
ultraviolet scale far beneath the Planck scale. For instance, we
claim that there must be a light charged particle with a small mass
\begin{equation}
m_{\rm el} \lsim g_{\rm el} M_{\rm Pl} \; .
\end{equation}
This statement should also hold for magnetic monopoles,
\begin{equation}
m_{\rm mag} \lsim g_{\rm mag} M_{\rm Pl} \sim \frac{1}{g_{\rm el}}
M_{\rm Pl}
\; .
\end{equation}
Note that the monopole masses are a probe of the ultraviolet cutoff
of a $U(1)$ gauge theory. The monopole has a mass at least of order
the energy stored in the magnetic field it generates; this is
linearly divergent, and if the theory has cutoff $\Lambda$, this is
of order
\begin{equation}
m_{\rm mag} \sim \frac{\Lambda}{g_{\rm el}^2} \;
\end{equation}
which is indeed parametrically correct in all familiar
examples---for the $U(1)$ arising in the Higgsing of an $SU(2)$,
this is the correct expression for the monopole mass with $\Lambda
\to m_W$, while on a lattice with spacing $a$, this is the monopole
mass with $\Lambda \to 1/a$. Therefore, the above constraint on
monopole masses tells us that for small $g$, the effective theory
must break down at a prematurely low scale
\begin{equation}
\Lambda \lsim g M_{\rm Pl} \; .
\end{equation}

This conjecture can be rephrased as the plausible statement that
``gravity is the weakest force''; if charged particles have $m > g
M_{\rm Pl}$, then the gauge repulsive force between them is overwhelmed
by the gravitational attraction---while if there are states with $m
\lsim g M_{\rm Pl}$, then gravity is subdominant. Of course, in highly
supersymmetric systems, all velocity-independent forces vanish in
any case, but as we will see, our constraints still apply.

Finally, we are familiar with restrictions on the relative
magnitudes of masses and charges from BPS bounds, but these have the
opposite sign, telling us that $M \geq Q$ in some units.
 Our bound might appear to be
 an ``{\it anti-}BPS'' bound.
This is not quite accurate.  First of all the BPS case
is a limiting case of our bound, saturating it.  Secondly
we are not conjecturing that for a given charge sector
all the masses are less than the charge, but that there exist
{\it some} such state.


We have phrased our constraint as one on the strength of $U(1)$
gauge couplings. This is a running coupling so we should ask about
the scale at which it should be evaluated. For the statement that
there exists a light charged particle with mass $m < g M_{Pl}$, it
is natural to use the asymptotic value of $g$,  the running coupling
evaluated at the mass of the lightest charged particle.  But for the
statement about the existence of a cutoff $\Lambda < g M_{Pl}$, it
is clearly most natural to consider the running coupling near the
scale $\Lambda$. Clearly our bound will also apply to non-Abelian
gauge theories, that can be Higgsed to $U(1)$'s. In this case, the
mass of the $W$'s is $m_W \sim g_{\rm el}v$ where $v$ is an
appropriate vev, and these particles will satisfy our bound as long
as the vevs don't exceed the Planck scale, a statement not
independent of conjecture about the finiteness of the volume of
moduli spaces in \cite{swamp}.

\subsection{Black holes and global symmetries}

Why should such a conjecture be true? One motivation has to do with
the well-known argument against the existence of global symmetries
in quantum gravity. Gauge symmetries are of course legal, but as we
take the limit $g \to 0$, the symmetry becomes physically
indistinguishable from a global symmetry. Something should stop this
from happening, and our conjecture provides an answer. As the gauge
coupling goes to zero $g \to 0$,  the cutoff on the effective theory
$\Lambda \to 0$ as well, so that the limit $g \to 0$ can not be
taken smoothly.

To see this more concretely, imagine that we have $g \sim
10^{-100}$, and consider a black hole with mass $\sim 10 M_{\rm Pl}$.
Since the gauge coupling is so tiny, the black hole can have any
charge between 0 and $\sim 10^{100}$, and still be consistent with
the bound for having a black hole solution ($M\geq QM_{\rm Pl}$). But if
there are no very light charged particles, none of this charge can
be radiated away as the black hole Hawking evaporates down to the
Planck scale. But then we will have a Planckian black hole labeled
by a charge anywhere from $0$ to $10^{100}$. This leads to
$10^{100}$ Planck scale remnants suffering from the same problems
that lead us to conclude quantum gravity shouldn't have global
symmetries (see e.g.\ \cite{remn}).

\OMITCOMPLETELY{
This heuristic argument would only lead to a logarithmic constraint
relating $\log(g)$ and the cutoff $\Lambda$. However, we may improve
the argument and obtain the same power law parametric dependence
that also follows from our conjecture. When the cutoff of the theory
is $\Lambda$, it is a good approximation to talk about black holes
whose radius is as small as $R=1/\Lambda$. The maximal
integer-valued charge of such a black hole is of order ${\cal N} =
M_{\rm Pl} / g\Lambda$ and it also determines a lower bound on the black
hole entropy because the black hole may be thought of as a bound
state of roughly ${\cal N}$ objects whose entropy $S$ is of order
one. Indeed, since we expect $S(N_1 + N_2) \geq S(N_1) + S(N_2)$, we
should have $S(N) \geq c N$ for some constant $c$. For some BPS
objects (like say D0-branes) the bound states of $N$ objects is
unique, so $c = 0$ and the inequality is saturated, but generically
it is reasonable to expect $c$ to be $\sim O(1)$. This entropy must
in turn be smaller than the Bekenstein-Hawking entropy $\sim R^2
M_{\rm Pl}^2 = M_{\rm Pl}^2/ \Lambda^2$ which implies our bound $$
\frac{M_{\rm Pl}}{g\Lambda} \lsim \frac{M_{\rm Pl}^2}{\Lambda^2}
\quad\Rightarrow \quad \Lambda\lsim g M_{\rm Pl} \; .$$
}

\vspace{4mm}

\qquad\qquad\qquad
\epsfig{file=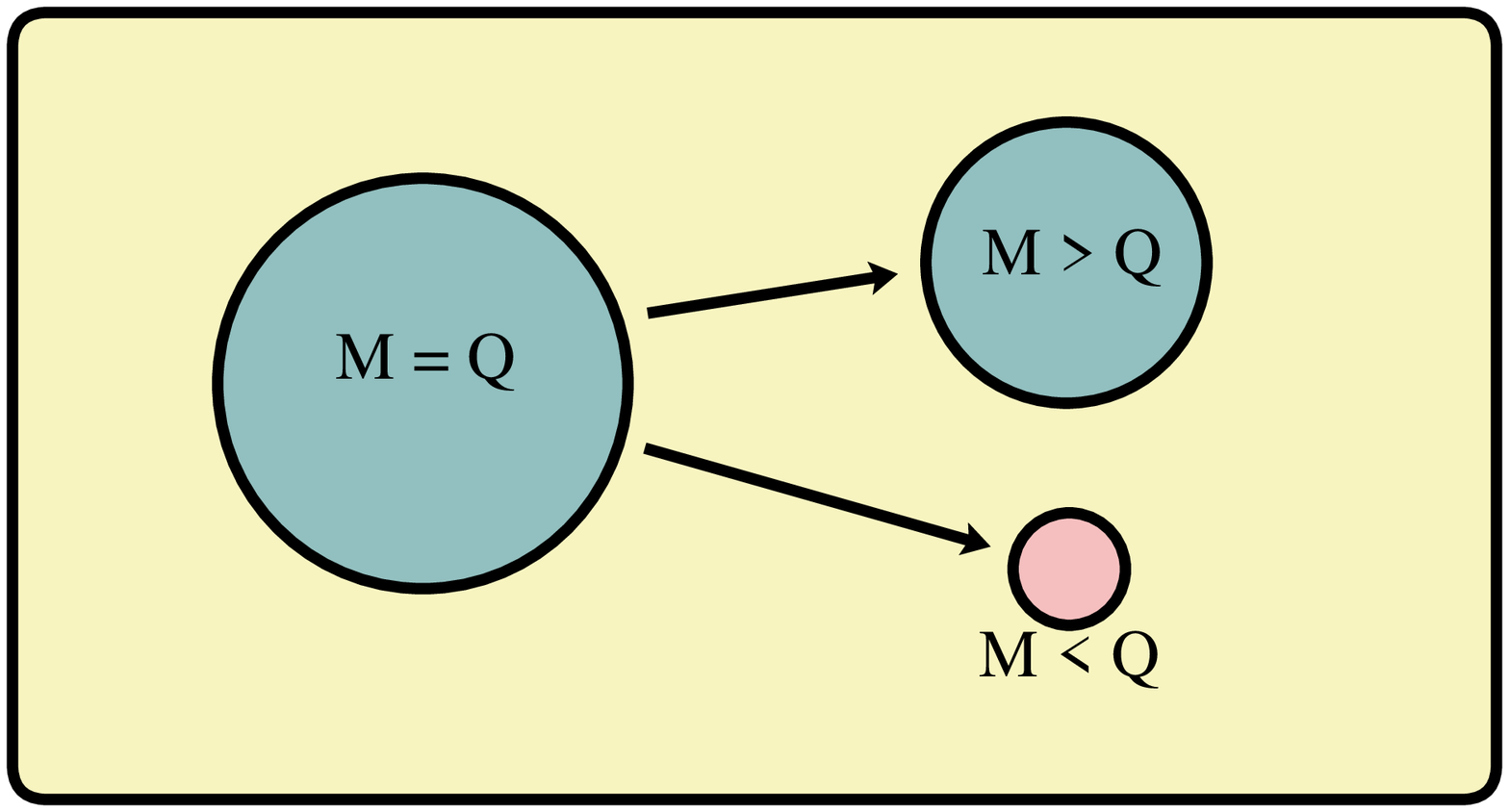,width=10cm}

\begin{quotation}{\bf Figure 2.} An extremal black hole can decay only
if there exist particles whose charge exceeds their mass.
\end{quotation}

The difficulties involving remnants are avoided if macroscopic black holes can
evaporate all their charge away, and so these states would not be
stable. Since extremal black holes have $M = Q M_{\rm Pl}$, in order for
them to be able to decay into elementary particles, these particles
should have $m < q M_{\rm Pl}$. Our conjecture also naturally follows
from Gell-Mann's totalitarian principle (``everything that is not
forbidden is compulsory'') because there should not exist a large
number of exactly stable objects (extremal black holes) whose
stability is not protected by any symmetries.

Another heuristic argument leading to same limit on $\Lambda$ is the
following. Consider the minimally charged monopole solution in the
theory. With a cutoff $\Lambda$, its mass is of order $M_{\rm mon} \sim
\Lambda/g^2$ and its size is of order $R_{\rm mon} \sim 1/\Lambda$. It
would be surprising for the {\it minimally} charged monopole to
already be a black hole because the values of all charges carried by a
black hole should be macroscopic (and effectively continuous); after all,
a black hole is a classical concept. Demanding that this monopole is not black
yields
\begin{equation}
\frac{M_{\rm mon}}{M_{\rm Pl}^2 R_{\rm mon}} \lsim 1
\qquad \Rightarrow \qquad \Lambda \lsim g
M_{\rm Pl}
\end{equation}

\subsection{Simple parametric checks}

It is easy to check the conjecture in a few familiar examples. For
$U(1)$'s coming from closed heterotic strings compactified to four
dimensions, for instance, we have
\begin{equation}
g M_{\rm Pl} \sim M_s \; ,
\end{equation}
and there are indeed light charged particles beneath the string
scale $M_s$, which also sets the cutoff of the effective theory. In
Kaluza-Klein theory, we have
\begin{equation}
g M_{\rm Pl} \sim \frac{1}{R} \; ;
\end{equation}
again there are charged particles at this scale and it also acts as
the cutoff of the low-energy 4D effective theory. By $T$-duality,
it is easy to see that the same thing will be true for winding
number gauge symmetries, too.

Next let's consider $U(1)$'s in type I string theories compactified
to 4D. Here
\begin{equation}
g M_{\rm Pl} \sim \frac{M_s}{\sqrt{g_s}}
\end{equation}
and indeed at weak string coupling this is even larger than the
string scale. At large coupling, we revert to a heterotic dual.
Similarly, we can consider a $U(1)$ living on a stack of
D$p$-branes. Then
\begin{equation}
g M_{\rm Pl} \sim \frac{M_s}{\sqrt{g_s (RM_s)^{(9 - p)}}}
\end{equation}
where $R$ is a typical radius of the space transverse to the brane.
Again, here we could try and violate the bound for $(R M_s) \ll 1$
but then we revert to a $T$-dual description.

If we move a single D-brane far away from others, the objects
charged under its $U(1)$ become long strings and can be made heavier
and heavier. For a non-compact space, it is clear that they can be
made arbitrarily heavy---but to have 4D gravity, the space must be
compact and this limits the mass of these charged particles. For a
brane of codimension bigger than two in the large dimensions, which
we'll take to have comparable sizes $R$, the back-reaction of the
brane on the geometry is small and we can limit $m_W \lsim M_s R$.
Then, for instance for D3-branes moving in a spacetime
with $n$ large dimensions we
have
\begin{equation}
\frac{m_W}{g M_{\rm Pl}} \sim \sqrt{\frac{g_s}{(M_s R)^{(n - 2)}}}
\end{equation}
so again for $n > 2$, $g_s < 1$, and $R > l_s$, the ratio is smaller
than one. The magnetic monopole is a long D-string attached to the
brane, so the analogous ratio depends on $1/g$ instead and
\begin{equation}
\frac{m_{\rm mon}}{(1/g) M_{\rm Pl}} \sim \sqrt{\frac{1}{(M_s R)^{(n -
2)}}}
\end{equation}
is again satisfied.

Things are parametrically marginal for $n=2$, but look like they
might be violated for $n = 1$. As an example, suppose we have an
$AdS_5 \times X$ space, with $L_{AdS}$ parametrically larger than
the string scale. By the introduction of a ``Planck brane'' we have a
warped compactification down to 4D. Suppose also that we can have
some D3-branes in the space with a $U(1)$ gauge field living on
them. To an effective field theorist, there is no obstacle to
imagining that the internal space $X$ is small, of order the string
scale. But this is in conflict with our conjecture. The $W$'s
charged under the $U(1)$ correspond to a long string ending on the
D3-brane and hence have a mass scaling as $L_{AdS}M_s^2$, the $U(1)$
gauge coupling is $g_{YM}^2 = g_s$ while $M_{\rm Pl}^2 = M_s^8 V_X
L_{AdS} / g_s^2$. The requirement that
\begin{equation}
m_W^2 \lsim g_{YM}^2 M_{\rm Pl}^2
\end{equation}
then implies that
\begin{equation}
(V_X M_s^5) \gsim g_s (L_{AdS} M_s) \; .
\end{equation}
In other words, the volume of the internal 5D space {\it must} be as
large as $g_s L_{AdS}$ in string units. This is certainly true for
the familiar $AdS_5 \times S^5$ compactifications, where the volume
of the $S^5$ scales like $L_{AdS}^5$. One might try to orbifold the
internal $S^5$ to smaller volumes, but there are only 3 $U(1)$'s one might
orbifold by, and since the size of the ``slices'' in the $S^5$ can't
get much smaller than $l_s$, this still assures that the volume of $X$
is larger than $L_{AdS}^2$ in Planck units and safely
satisfies our bound. We aren't aware of any
examples where $X$ can be kept at the string scale with
parametrically large $L_{AdS}$.

Finally, let us consider another possible way in which our bound
might be parametrically violated. Consider a theory with some cutoff
$\Lambda$ but  a large $N$ number of $U(1)$'s, perhaps associated
with wrapped brane charges along a large number of cycles in some
compactification. If we could somehow Higgs these $U(1)$'s down to
the diagonal subgroup, the low-energy coupling $g_{diag}$ would be
suppressed by $\sim \frac{1}{\sqrt{N}}$, and we can make the
coupling very weak. However, at large $N$, there is also a ``species
problem''---$N$ can't be made parametrically large without making
gravity weak \cite{remn}.
For instance, naively the cycles would have to each
occupy a volume of order the string scale, so that the internal
volume and hence $M_{\rm Pl}^2$ also grows as $N$, and hence $N$ cancels
out of the combination $g_{diag} M_{\rm Pl}$. Similar issues were
discussed in the ``$N$-flation'' proposal of \cite{Nflate}.

\subsection{Generalizations}

It is natural to generalize our loose conjecture: consider a
$p$-form Abelian gauge field in any number of dimensions $D$; then
there are electrically and magnetically charged $p - 1$ and $D - p -
1$ dimensional objects with tensions
\begin{equation}
T_{\rm el} \lsim \Big(\frac{g^2}{G_{\rm N}}\Big)^{1/2}, \qquad
T_{\rm mag} \lsim \Big(\frac{1}{g^2
G_{\rm N}}\Big)^{1/2} ,
\end{equation}
where the coupling $g$ (the charge density) has a dimension of
$\,  mass^{\,p+1-D/2}$.
We have discussed the case $D= 4, \,\, p =1$ above.
Whenever the objects carry central charges only, the inequalities above
are saturated and coincide with the BPS bounds. However, the inequality
becomes strict for other types of charges. This generalization
can be used to rule out effective field theories that have been
constructed in the literature for a variety of purposes.

\vspace{4mm}

\qquad\qquad\qquad\qquad
\epsfig{file=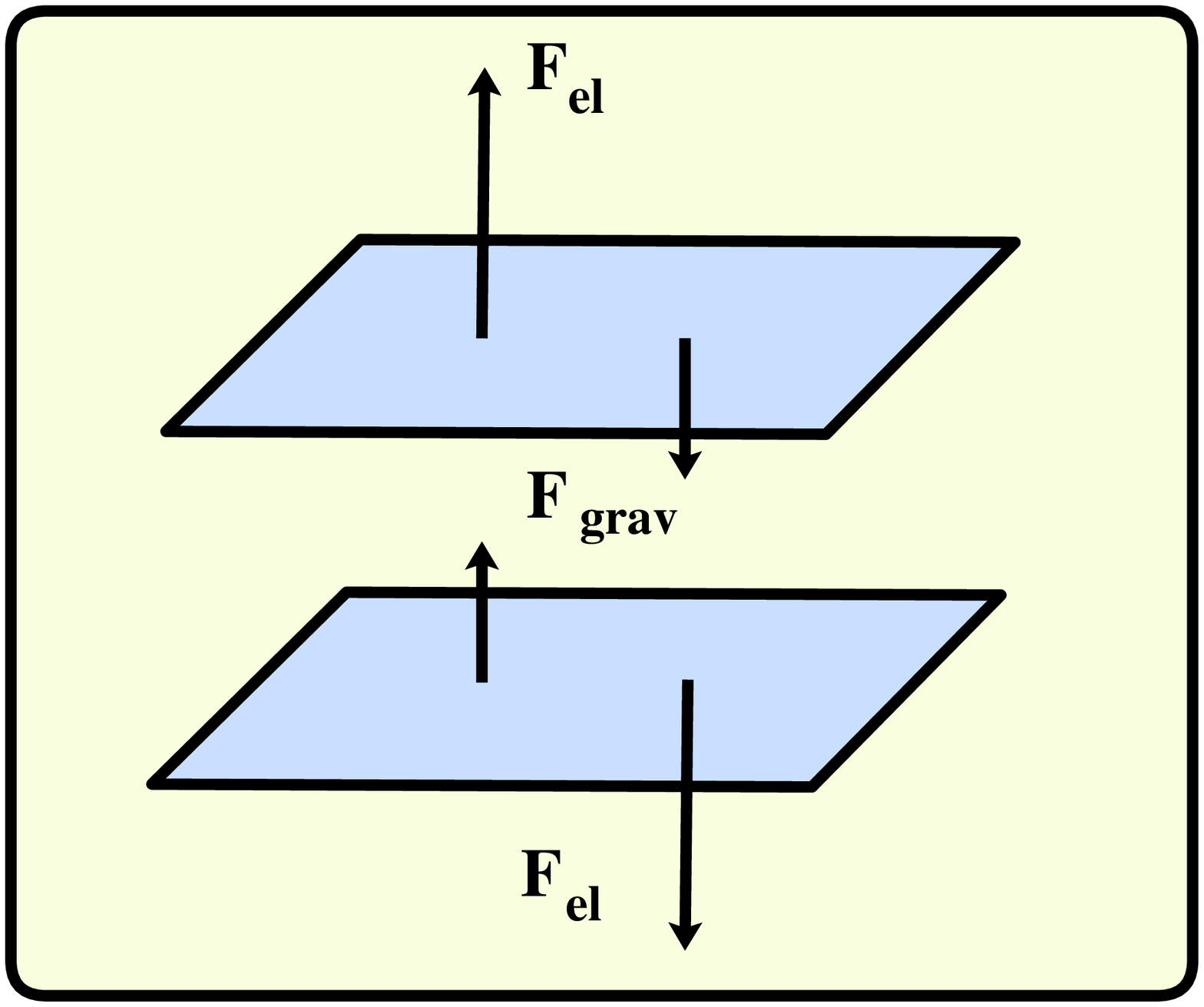,width=8cm}

\begin{quotation}{\bf Figure 3.}
For any kind of an electric field, there should exist ``self-non-attractive''
objects for whom the electric repulsive force exceeds the strength of gravity.
\end{quotation}

For instance, in \cite{infl},
it was argued that a natural candidate
for an inflaton could be found in 5D gauge theories compactified on
a circle. Consider a $U(1)$ gauge theory with gauge coupling $g_5$
and Planck scale $M_5^3$. Compactifying on a circle, we have 4D
gravity as well as the 4D periodic scalar $\theta = A_5 R$
associated with the Wilson line around the circle. The effective
action for $\theta$ and gravity is simply
\begin{equation}
\int d^4 x \sqrt{-g} \left[F^2 (\partial \theta)^2 + M_{\rm Pl}^2 R
\right] ,\end{equation} where
\begin{equation}
F^2 = \frac{1}{g_5^2 R} \; , \qquad
M_{\rm Pl}^2 = M_5^3 R \; .
\end{equation}
At one-loop level,
light charged scalars can generate a potential $V(\theta)
\sim (1/R^4) \cos(\theta)$. It is easy to see that $\theta$ can have
slow-roll inflation as long as $F^2 \gg M_{\rm Pl}^2$, or
\begin{equation}
g_5^2 M_5^3 R^2 \ll 1 \; .
\end{equation}
This is perfectly consistent in effective field theory, but it runs
afoul of our general constraint for $D=5,\,\, p = 1$. Indeed,
straightforward attempts to embed this model into string theory
fail. For instance, for the $U(1)$ coming from closed strings,
$g_5^2 M_5^3 = M_s^2$ and so we must have $R \ll M_s^{-1}$, where
this description breaks down and the $T$-dual is appropriate.

Indeed, Banks, Dine, Fox, and Gorbatov \cite{bdine} subsequently
studied the more general question of whether in compactifications to
4D it is possible to get periodic scalars (``axions'') with decay
constants $F$ parametrically larger than the Planck scale. In all
the examples they studied, they found that either $F$ can't be made
larger than $M_{\rm Pl}$, or that if it could, there was also an
instanton of anomalously small action $S_{\rm inst} \sim M_{\rm Pl}/F$, so
the instanton generated unsuppressed potential generated terms up to
$\cos (N \theta)$ with $N \sim M_{\rm Pl}/F$, ruining the parametric
flatness of the potential. This observation is subsumed in our
generalized conjecture; for $D=4, \,\, p = 0$, the 0-form is an
axion; the ``tension'' of the object charged under it is simply the
action of an instanton coupling to the axion, while the axion gauge
coupling is $\,g \sim 1/F\,$ where $F$ is the axion decay constant
so our constraint gives precisely
\begin{equation}
S_{\rm inst} \lsim \frac{M_{\rm Pl}}{F} \; .
\end{equation}

\section{Sharpening the claim}

Working in $M_{\rm Pl}$ = 1 units, we are making a conjecture about
mass/charge ratios
\begin{equation}
(M/Q) \lsim 1
\end{equation}
To find a sharper conjecture, we have to decide (a) what states
should satisfy this bound and (b) what to mean by ``1''. For the last
point, it is natural to take the $(M/Q)$ ratio that is equal to one
for large extremal black
holes. As for the states to consider, there are three natural
possibilities:
\begin{eqnarray}
\mbox{(I)} \, \, \left(\frac{M_{q_{\rm min}}}{q_{\rm min}}\right) &\leq& 1,
\qquad\mbox{for the
state of minimal charge;} \nonumber \\
\mbox{(II)\,}  \, \left(\frac{M_{\rm min}}{q_{M_{\rm min}}} \right)&\leq& 1,
\qquad\mbox{for the lightest charged particle;} \nonumber \\
\mbox{(III)\,}  \, \left(\frac{M}{q}\right)_{\rm min} &\leq& 1,
\qquad\mbox{for the state with smallest mass/charge ratio.} \nonumber
\end{eqnarray}

Of course, for these statements to have a sharp meaning, the state
must be exactly stable for $M$ to be meaningful. The particle of
smallest charge is {\it not} guaranteed to be stable---for instance,
a heavy charged particle of charge $+1$ can decay into two lighter
charged particles of charge $-2,+3$. If the particles with charges $-2$ and
$+3$ are light, they will form a Kepler/Coulomb bound state of charge $+1$.
This
state will be stable but its $M/Q$ ratio may be larger than for the
states with charges $-2$ and $+3$. In particular, it may be larger than
one.

Furthermore, there are easy
counterexamples to the conjecture (I) in string theory, even when the
minimally charged particles {\it are} exactly stable. For instance,
in the weakly coupled $SO(32)$ heterotic string, the spinor of
$SO(32)$ is exactly stable and has minimal half-integral charges
under the $U(1)$'s inside the $SO(32)$, but is heavy and can violate
our bounds. A generalization of these are the half-integrally
charged winding strings considered by Wen and Witten \cite{wenwitten},
that have
fractionally charges but are also heavy. So (I) can't be right.

Of course the {\it lightest} charged particle is exactly stable, as
is the particle with smallest $(M/Q)$ (as follows trivially from the
triangle inequality), so both (II) and (III) are well-defined
conjectures. Obviously (II) is the stronger of the two (and it
clearly implies (III)) and {\it
forces} the effective theory to contain a light charged particle.
Conjecture (III) can in principle be satisfied by a heavy state with
large $Q$ which would reduce the impact of the inequality on physics
at low energies.  Even though we have no counterexamples for conjecture
(II) most of our evidence only supports the weaker conjecture (III).

When there are several $U(1)$'s, the generalization of the
conjecture is clear. In every direction in charge space, including
electric and magnetic charges, at large values of the charges, we
have extremal black hole solutions. The conjectures (II) and (III)
then imply the existence of light charged particles with $(M/Q) <
(M/Q)_{\rm extremal}$ in certain directions of charge space. More
precisely, there should always exist a set of directions in the
charge space that form a basis of the full space where the
inequality is satisfied.

It is interesting to see what the spectrum of a theory which
violates our conjectures looks like. Suppose for simplicity that
there is only one ``elementary'' particle with minimal charge $1$,
but with $M > Q$. Since the net force between two of these particles
is attractive, there is a Kepler bound state of two of these
particles, with charge $2$, but with a mass smaller than $2 M$, so
that the mass/charge ratio decreases. We can continue to add further
particles to make further bound states, with $(M/Q)$ continually
decreasing. This proceeds till the bound state eventually turns into
an extremal black holes, and asymptotically, we reach $(M/Q) = 1$.
It is easy to see that {\it all} of these particles are exactly
stable: since $(M/Q)$ is a decreasing function of $Q$, none of these
states can decay into a collection of particles with smaller
charges.

On the other hand, if there are any states with $(M/Q) < 1$, then
the macroscopic black holes can always decay, and the number of
exactly stable particles will be finite. Suppose that, among the
states with $(M/Q) < 1$, the one with smallest charge has charge
$Q_{\rm min}$. Then, by the same argument as above, we expect that
the lightest particles with charges smaller than $Q_{\rm min}$ are
exactly stable.

So our mass/charge ratio conjecture (III) can be seen to follow from
a very simple general conjecture valid for both charged and
uncharged particles: The number of exactly stable particles in a
theory of quantum gravity in asymptotically flat space is finite.
Actually this statement is not quite correct. Clearly we can have an
infinite number of exactly stable BPS states, and many of these are
safely bound; consider for instance dyons of electric/magnetic
charge $(n,1)$ for large $n$. However, the number of exactly stable
(and safely bound) states {\it in any given direction in charge
space} is finite.

Even for neutral particles, this implies that the number of massless
degrees of freedom is finite, and such a restriction is indeed
suggested by the species problem associated with the Bekenstein
bound. If there is a principle dictating the number of exactly
stable particles to be finite, it is reasonable to expect that in
all the vacua in the landscape, the number of exactly stable states
is typically of order a few. In this case, the minimal charge
$Q_{\rm min}$ for which $(M/Q) < 1$ should not be too large, since
as we saw above the number of exactly stable states grows with
$Q_{\rm min}$. This then substantiates our loose conjectures $m
\lsim g M_{\rm Pl}$.

\section{Evidence for the conjecture}

Our conjecture is now phrased sharply enough that we can look for
non-trivial checks of it in known stringy backgrounds. Clearly in
highly supersymmetric situations where $U(1)$'s are associated with
central charges, there will be BPS states saturating our
inequality. This will for instance be the case in theories with 32
supercharges. However, already with 16 supercharges non-trivial
checks are possible, for instance in compactifications of the
heterotic string on tori with generic Wilson lines, where most of
the $U(1)$'s are not central charges.

Consider for instance the $SO(32)$ heterotic string compactified on
$T^6$. At a generic point on moduli space, there is a $U(1)^{28}$
gauge symmetry. We will check our conjecture for electric charges
only; by $S$-duality, this check will carry over to magnetic charges
as well. A general set of electric charges is a 28-dimensional
vector
\begin{equation}
Q = \left(\frac{Q_L}{Q_R}\right)
\end{equation}
where $Q_L$ is 22-dimensional vector and $Q_R$ is 6-dimensional
vector. The charges are quantized, lying on the 28-dimensional even
self-dual lattice with
\begin{equation}
Q_L^2 - Q_R^2 \in 2 {\mathbbm Z}
\end{equation}
Moving around in moduli space corresponds to making $SO(22,6)$
Lorentz transformations on the charges.

\vspace{4mm}

\qquad\qquad\qquad\qquad
\epsfig{file=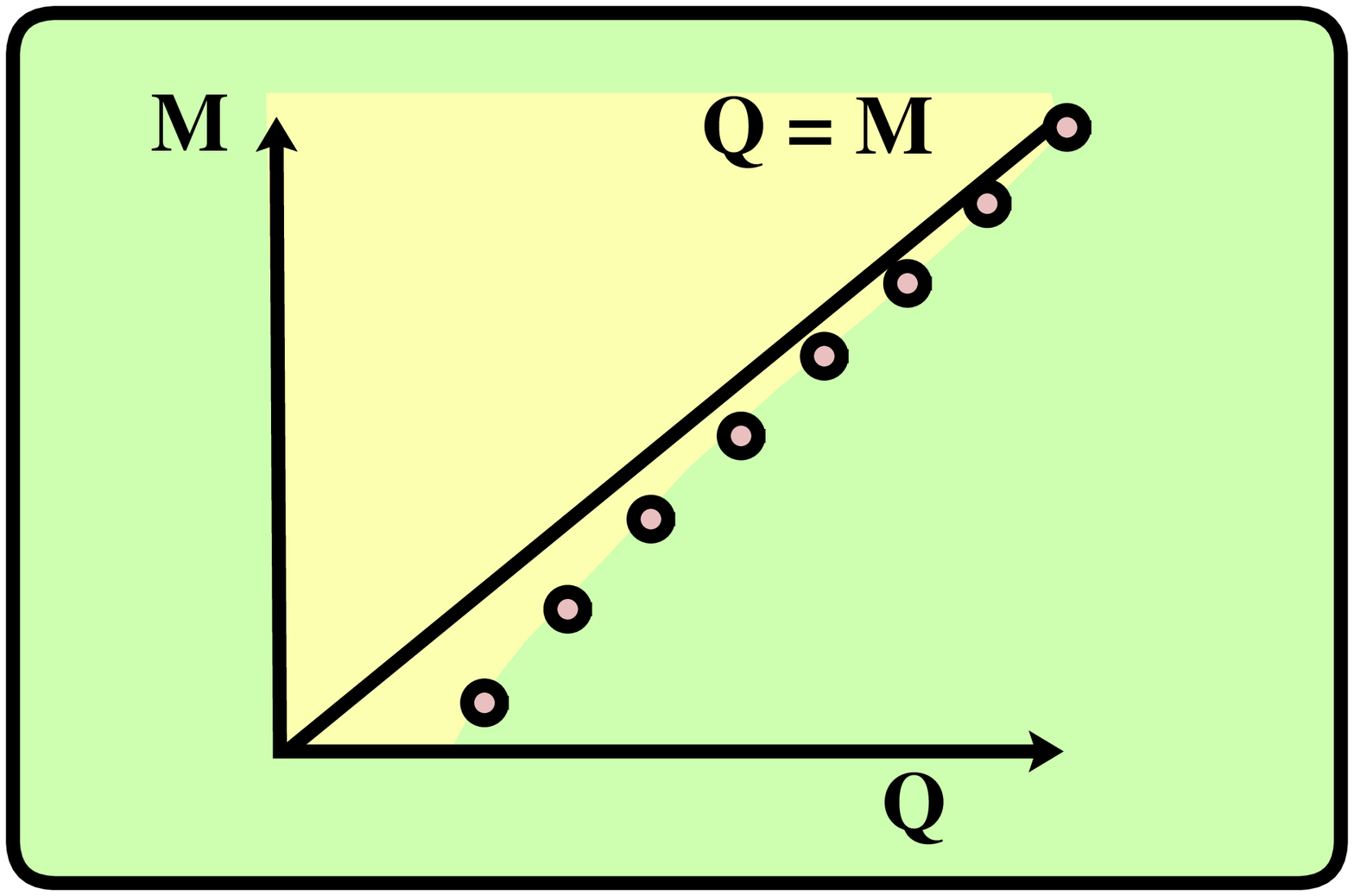,width=8cm}

\begin{quotation}{\bf Figure 4.} The charge $M$ of the heterotic
string states of charge $Q$ approaches the $M=Q$ line from below.
The yellow area denotes the allowed region.
\end{quotation}

The extremal black hole solutions in this theory were constructed by
Sen \cite{sen}. For $Q_R^2 - Q_L^2 > 0$, there are BPS black hole solutions
with mass
\begin{equation}
M^2 = \frac{1}{2} Q_R^2
\end{equation}
where we work in units with $M_{\rm Pl} = 1$. For $Q_L^2 - Q_R^2 > 0$,
the black holes are {\it not} BPS; still, the extremal black holes
have mass
\begin{equation}
M^2 = \frac{1}{2} Q_L^2 \; .
\end{equation}
We can compare this with the spectrum of perturbative heterotic
string states, given by
\begin{equation}
M^2 = \frac{1}{2} Q_R^2 + N_R = \frac{1}{2} Q_L^2 + N_L - 1
\end{equation}
where $N_{R,L}$ are the string oscillator contributions
and where we chose units with $\alpha' = 4$. The $-1$,
coming from the tachyon in the left-moving bosonic string, is
crucial. Note that this spectrum nicely explains the BH spectrum of
the theory, as the highly excited strings are progenitors of
extremal black holes. Consider large $\,Q_L,Q_R\,$, with $Q_R^2 >
Q_L^2$. Then, the minimal $M^2$ compatible with these charges will
have $N_R = 0,\,\, N_L = \frac{1}{2} (Q_R^2 - Q_L^2) + 1$, which are
BPS, with $M^2 = \frac{1}{2} Q_R^2$. On the other hand, for $Q_L^2 >
Q_R^2$, the minimal $M^2$ is with $N_L = 0$, and $N_R =
\frac{1}{2}(Q_L^2 - Q_R^2) - 1$. These are not BPS, but for large
$Q_L^2$, they have $M^2 = \frac 12 Q_L^2 $.

But the string spectrum also guarantees that, as we go down to
smaller charges along a basis of directions in charge space, we are
guaranteed to find a state with a mass/charge ratio smaller than for
extremal BH's. The inequality is saturated for the BPS states which
have $Q_R^2 > Q_L^2$, but for $Q_L^2 > Q_R^2$ the extremal black
holes have $M^2 = \frac 12 Q_L^2 $ while there is always a state with
mass
\begin{equation} M^2 = \frac{1}{2} Q_R^2= \frac{1}{2} Q_L^2 - 1
\end{equation}
since there is a charge vector with $Q_L^2 - Q_R^2 = 2$ on the
charge lattice.


\subsection{Gauge symmetries vs. global symmetries}

It is possible to generalize this argument to {\it any} perturbative
heterotic string compactification, including
compactifications on $K3$ and arbitrary Calabi-Yau threefolds,
as a straightforward generalization of the
familiar argument that all global symmetries in this theory are
gauged. For any integral $U(1)$ gauge symmetry coming from
the left-movers of heterotic string, there is a worldsheet current
$J(z)$.
Because it is a $(1,0)$ primary field,
one can construct the $(1,1)$ vertex operator
$J(z) \,\bar \partial X_\mu \,  e^{i k X}(z,\bar z)$ with $k^2 = 0$
for a spacetime gauge field and prove that the corresponding
symmetry is a gauge symmetry. (Analogously, a symmetry
coming from the right-movers would be associated with a current
$\tilde J(\bar z)$, a $(0,1)$ primary field, and the vertex
operator would be $\tilde J(\bar z)\,\partial X_\mu\,
e^{i k X}(z,\bar z)$ with $k^2=0$.)
We can take
the worldsheet CFT to consist of the $U(1)$ part together with the
rest. We can bosonize the current with level $k$ as
\begin{equation}
J(z) \sim \sqrt{k}\,  \partial \phi(z) \; .
\end{equation}
Then, the operator
\begin{equation}
O(z) = \,: \!e^{i \sqrt{k} \phi(z)}\!:
\end{equation}
is always in the CFT.  There are a number of ways
to see this.  One way is to note that this operator
has local OPE with all the operators in the theory.  In fact
using the integrality of the $U(1)$ charge, the
content of any operator will be of the form
$$V\sim :\exp (i p \phi / \sqrt{k}):\cdot V'$$
where $p$ is an integer-valued charge and
where $V'$ has no exponential parts in $\phi$.
It is easy to see that $O(z)$ will have local OPE with this operator.
Therefore, the completeness of the CFT spectrum (i.e.~the statement
that the operator content of the theory is maximal consistent
with local OPE, as follows from modular invariance) forces us to have
$O(z)$ as an allowed operator in the theory.

Another way to understand the existence of the operator $O(z)$
is to note that it corresponds to spectral flow by 1 unit in the $U(1)$.
This simply corresponds to changing the boundary conditions on the circle
by $\exp(2\pi i \, \theta \, p)$ where $p$ denotes the $U(1)$ charge and
$\theta$ goes from $0$ to $1$.

We thus see that the state corresponding to $O(z)$ exists
in the spectrum of CFT. Since by assumption
this is a left-mover state, this corresponds to $N_L =
0$ and so $M^2 = \frac{1}{2} Q_L^2 - 1$, while asymptotically, the
excited strings correspond to extremal black holes with $M^2 =
\frac{1}{2} Q_L^2$, so our string state is indeed {\it
sub}-extremal.

\section{Possible relation to subluminal positivity constraints}

It is natural to conjecture that since there must exist states for
which $(M/Q) < 1$ while the extremal black holes have $(M/Q) = 1$,
the extremal limit for $(M/Q)$ for black holes is approached from
below, that is, that the leading corrections to the extremal black
hole masses from higher-dimension operators should again {\it
decrease} the mass. This implies some positivity constraint on some
combination of higher-dimension operators.

It is interesting that similar positivity constraints have been
discussed in \cite{superlum}, where it was found that certain
higher-dimension operators must have positive coefficients in order
to avoid the related diseases of superluminal signal propagation
around configurations with a nonzero field strength
and bad analytic properties of the $S$-matrix. For instance,
consider the theory of a $U(1)$ gauge field in four dimensions. The
leading interactions are $F^4$ terms, and the effective Lagrangian
is of the form
\begin{equation}
-
F_{\mu \nu}^2 + a (F^2)^2 + b (F \tilde F)^2 + \cdots
\label{feight}
\end{equation}
If the scale
suppressing the dimension 8 operators
is far beneath the Planck scale, we can ignore
gravity, and the claim of \cite{superlum} is that $a,b$ must be
positive to avoid superluminal propagation of signals around
backgrounds with uniform electric or magnetic fields, and also to
satisfy analyticity and dispersion relation for the photon-photon
scattering amplitude.

Of course these higher dimension operators also change the
mass/charge relation for extremal black holes. Indeed, there are many
other operators which do this as well; at the leading order they include
$R^2$ and $R F^2$ type terms as well. But we can imagine that the
$F^4$ terms dominate in the limit where
the scale suppressing the $F^4$ terms
is far smaller than the
Planck scale. Treating the $a,b$ terms as perturbations, we can
solve for the modified Black Hole background to first order in
$a,b$, and find the new bound on $M$ which has a horizon and no
naked singularity. To first order in $a,b$, we find that for a black
hole with electric and magnetic charges $(Q_e,Q_m)$ and working with
$M_{\rm Pl} = 1$
\begin{equation}
M^2_{\rm extr} = (Q_e^2 + Q_m^2)
- \frac{2 a}{5} \frac{(Q_e^2 - Q_m^2)^2}{(Q_e^2 + Q_m^2)^2}
- \frac{32 b}{5} \frac{Q_e^2 Q_m^2}{(Q_e^2 + Q_m^2)^2}\label{elmgbh}
\end{equation}
So, for purely electric or magnetic black holes, we have
\begin{equation}
M^2_{\rm extr} = Q^2 - \frac{2 a}{5}
\end{equation}
which indeed decreases for the ``right'' sign of $a > 0$.
The same statement holds for the dyonic black holes as long as $b>0$
which is also the ``right'' sign.
The result \eqref{elmgbh} has, in fact, an $SO(2)$ symmetry
mixing $Q_e$ and $Q_m$ for $a=4b$, much like
the stress-energy tensor derived from \eqref{feight}
for the same values $a=4b$.\footnote{While
our inequality $M_{\rm extr}<|Q|$ holds uniformly for $a>0,b>0$, we would
also be able to satisfy our constraint and
find a basis of directions where the inequality holds
whenever at least one parameter ($a$ or $b$) is positive.}
 The effect of other four-derivative terms on the extremal black hole
masses will be studied elsewhere \cite{padi}.

There is another hint of a connection between our work and
\cite{superlum}. The superluminality/analyticity constraints were
shown to be violated by the Dvali-Gabadadze-Porrati \cite{dvaligp}
brane-world model for modifying gravity in the IR. Interestingly,
this model represents another example of trying to make interactions
in the theory much weaker than gravity: the model has a 5D bulk with
Planck scale $M_5$, but with a large induced Einstein-Hilbert action
$\int d^4 x \sqrt{-g_{\rm ind}}\, M_{\rm Pl}^2 R^{(4)}$ on the brane. With
$M_{\rm Pl} \gg M_5$, this (quasi)-localizes gravity on the 4D brane.
Again naively, there is nothing wrong with taking $M_{\rm Pl}$ large, as
it seems to make the theory more weakly coupled; in this way it is
similar to taking the limit of tiny gauge couplings in our
examples, but we can here prove that the theory leads to
superluminality and acausality in the IR, and is inconsistent with
the standard analyticity properties of the $S$-matrix.

\section{Discussion}
In this note, we have argued that there is a simple but powerful
constraint on low-energy effective theories containing gravity and
$U(1)$ gauge fields. An effective field theorist would not see any
problem with an arbitrarily weak gauge coupling $g$, but we have
argued that in fact there is a hidden ultraviolet scale $\Lambda
\sim g M_{\rm Pl}$, where the effective field theory breaks down, and
that there are light charged particles with mass smaller than
$\Lambda$. While this statement is completely unexpected to an
effective field theorist, it resonates nicely with the impossibility
of having global symmetries in quantum gravity, and the associated
ability for large charged black holes to dissipate their charge in
evaporating down to the Planck scale.

The specific forms of our conjecture are sharp, and if they are
wrong it should be possible to find simple counter-examples in
string theory, though we have not found any. The strongest form is
that for the lightest charged particle along the direction of some
basis vectors in charge space, the $(M/Q)$ ratio is smaller than for
extremal black holes. Such an assumption allows
all extremal black holes to decay
into these states. The weaker statement says that there should exist
{\it some} state with mass/charge ratio smaller than for extremal
black holes. In all the examples we have seen, this state has a
``reasonably small'' charge, so it is light; however, the weaker form
allows the possibility that the smallest $M/Q$ is realized for some
large charge $Q_*$ and objects that are ``nearly'' extremal black
holes. While the number of exactly stable states would be finite in
this case, it would still be extremely large. If this weaker form of
the conjecture is true it is likely that there is some distribution
of $Q_*$ peaked for charges of order 1, but perhaps with sporadic
exceptions at larger $Q_*$.

\vspace{4mm}

\qquad\qquad\quad
\epsfig{file=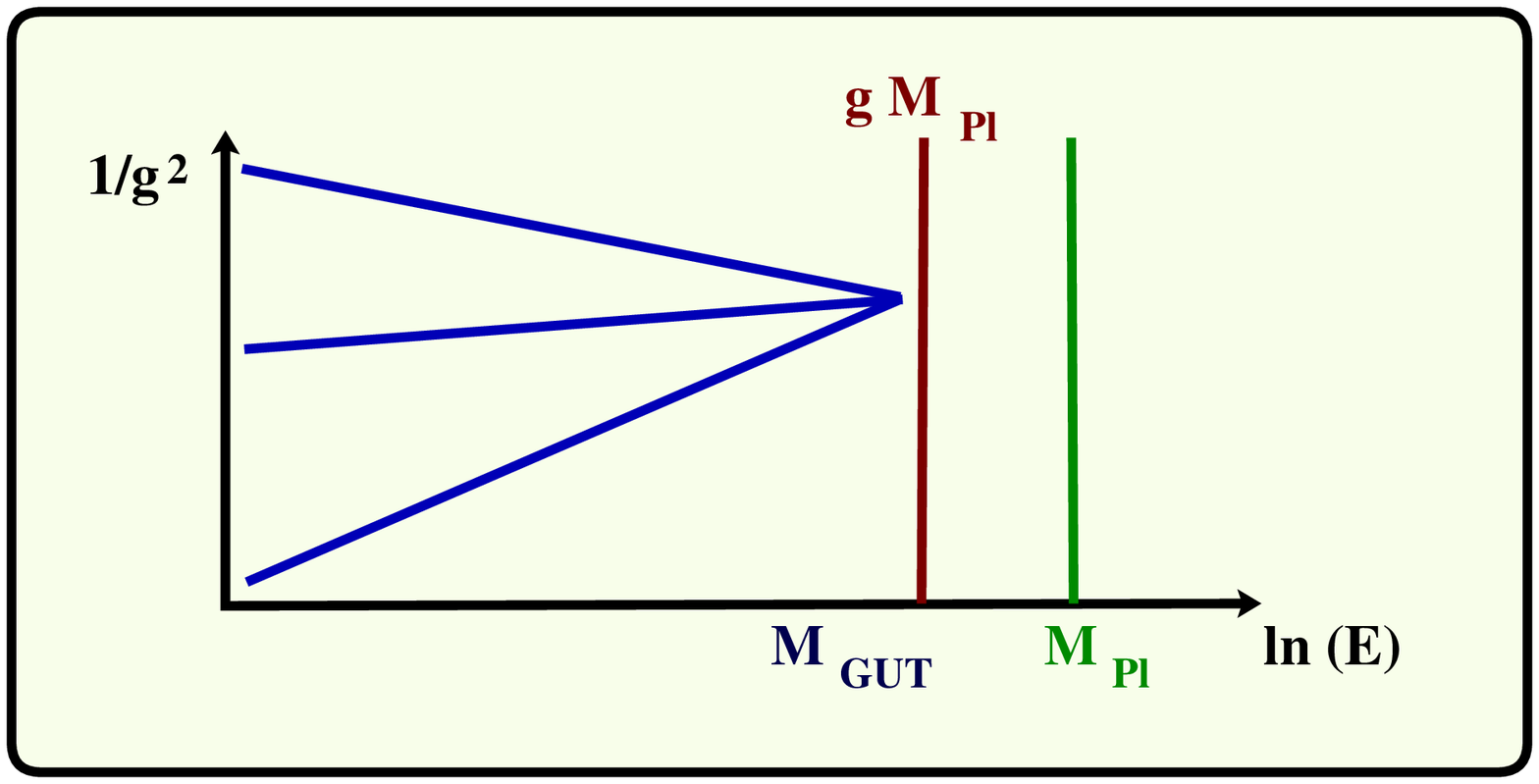,width=11cm}

\begin{quotation}{\bf Figure 5.}
Because the gauge couplings at very high scales are smaller than one,
our conjecture naturally predicts the existence of a new scale beneath
the Planck scale.
\end{quotation}

If true, our conjecture shows that gravity and the other gauge
forces can not be treated independently. In particular, any approach
to quantum gravity that begins by treating pure gravity and is able
to add arbitrary low-energy field content with any interactions is
clearly excluded by our conjecture. Of course in string theory all
the interactions are unified in a way that makes treating them
separately impossible. In particular, if we take the standard model
gauge (augmented by SUSY or split SUSY or other particles leading to
precision gauge coupling unification), we have perturbative gauge
couplings at a very high energy scale, and our conjecture then
implies that there {\it must} be new physics at a scale beneath the
Planck scale, given by $\Lambda \sim \sqrt{\alpha/G_{\rm N}}$ which is
close to the familiar heterotic string scale $\sim 10^{17}$ GeV.

Our conjecture also offers a new experimental handle on ultraviolet
physics, by searching for extremely weak new gauge forces. Indeed,
if a new gauge force coupling to, say, $B-L$ is discovered, with coupling
$g \sim 10^{-15}$, in the current generation sub-millimeter
force experiments, we would claim that there must be new physics at
an ultraviolet scale $\sim g M_{\rm Pl} \sim$~TeV. Forces of this
strength naturally arise in the context of large extra dimensions
with fundamental scale near a TeV \cite{add3}; what is interesting
is our claim that new physics {\it must} show up near the TeV scale.

It would be interesting to investigate whether there is an analogous
conjecture in Anti-de-Sitter spaces, since here it can be translated
into a statement about the spectrum of operators in the dual CFT
that can perhaps be proved on general grounds.

Finally, it is interesting that the constraint implied by our
conjecture seems to at least parametrically exclude apparently
natural models for inflation based on periodic scalars with
super-Planckian decay constants, which seem perfectly sensible from
the point of view of a consistent effective theory. Of course, in
the real world we don't need a parametrically large decay constant
to get parametrically large numbers of $e$-foldings of
inflation---60 $e$-foldings will do. If the strong form of our
conjecture is true, one might be tempted to conclude that there is a
sharp obstacle to getting this sort of inflation in quantum gravity.
If as is more likely the weaker form is true, then one might say
that even though the low-energy theorists' notion of technical
naturalness is misleading and such models are non-generic, there
might be sporadic examples where they are possible. Clearly these
are two very different pictures. The latter is more consistent with
much of the philosophy of exploration in the landscape so far:
things like a small cosmological constant are taken to be
non-generic, tuned,  but possible. But it is extremely interesting
that phenomena of clear physical interest, like inflation with
trans-Planckian excursions for the inflaton, which might even be
forced on us experimentally by the discovery of primordial
gravitational waves, seem to be pushing up against the limits of
what quantum gravity seems to want to allow. Further exploration of
the boundaries between the swampland and the landscape should shed
more light on these issues.

\section{Acknowledgements}

We are grateful to Shamit Kachru, Megha Padi, and Joe
Polchinski for discussions. We thank Jacques Distler for a
discussion on the scale-dependence of our loose conjectures, and
Juan Maldacena for clarifying discussion on the number of exactly
stable states in quantum gravity. The work of NAH is supported by
the DOE under contract DE-FG02-91ER40654. The work of LM is
supported by a DOE OJI award. CV is supported in part by NSF grants
PHY-0244821 and DMS-0244464. 


\newpage

\begin{thebibliography}{99}

\bibitem{swamp}
  C.~Vafa,
  {\it ``The string landscape and the swampland,''}
  arXiv:hep-th/0509212.

\bibitem{doug}
M.~Douglas, talk presented at Strings 2005, Toronto, Canada.

\bibitem{bdine}
  T.~Banks, M.~Dine, P.~J.~Fox and E.~Gorbatov,
  {\it ``On the possibility of large axion decay constants,''}
  JCAP {\bf 0306}, 001 (2003)
  [arXiv:hep-th/0303252].

\bibitem{remn}
L.~Susskind,
{\it ``Trouble for remnants,''}
arXiv:hep-th/9501106.

\bibitem{Nflate}
  S.~Dimopoulos, S.~Kachru, J.~McGreevy and J.~G.~Wacker,
  {\it ``N-flation,''}
  arXiv:hep-th/0507205.

\bibitem{infl}
  N.~Arkani-Hamed, H.~C.~Cheng, P.~Creminelli and L.~Randall,
  {\it ``Extranatural inflation,''}
  Phys.\ Rev.\ Lett.\  {\bf 90}, 221302 (2003)
  [arXiv:hep-th/0301218].

\bibitem{wenwitten}
  X.~G.~Wen and E.~Witten,
  {\it ``Electric And Magnetic Charges In Superstring Models,''}
  Nucl.\ Phys.\ B {\bf 261}, 651 (1985).

\bibitem{sen}
  A.~Sen,
  {\it ``Black hole solutions in heterotic string theory on a torus,''}
  Nucl.\ Phys.\ B {\bf 440}, 421 (1995)
  [arXiv:hep-th/9411187].

\bibitem{superlum}
    A.~Adams, N.~Arkani-Hamed, S.~Dubovsky, A.~Nicolis and R.~Rattazzi,
    {\it ``Causality, Analyticity and an IR Obstruction to UV Completion,''}
arXiv:hep-th/0602178.


\bibitem{padi}
    M.~Padi, L.~Motl et al., work in progress.


\bibitem{dvaligp}
G.~R.~Dvali, G.~Gabadadze and M.~Porrati,
{\it ``4D gravity on a brane in 5D Minkowski space,''}
Phys.\ Lett.\ B {\bf 485}, 208 (2000)
[arXiv:hep-th/0005016].

\bibitem{add3}
N.~Arkani-Hamed, S.~Dimopoulos and G.~R.~Dvali,
{\it ``Phenomenology, astrophysics and cosmology of theories with  sub-millimeter
dimensions and TeV scale quantum gravity,''}
Phys.\ Rev.\ D {\bf 59}, 086004 (1999)
[arXiv:hep-ph/9807344].



\end{thebibliography}
\end{document}